\documentstyle[epsf,12pt]{article}
\parskip=\medskipamount
\begin{document}

\begin{center}
{\large\bf Approximating Radiative Corrections
for Analytical Applications in the Data Analysis}\\[1cm]

{\large G.I. Smirnov}\\[0.8cm]

Laboratoire de Physique Corpusculaire 
de Clermont-Ferrand IN2P3/CNRS and  Univesit\'{e} Blaise Pascal
Clermont II, France 

and

Laboratory of Particle Physics, JINR, Dubna, Russia. 

\end{center}

\begin{abstract}
Exact calculations of the radiative corrections to the single
spin asymmetry in the exclusive pion electroproduction below
$\Delta(1232)$ resonance are performed by employing
the  {\small EXCLURAD} code. The analysis of the dependence
of the obtained results on $W$, $\theta_{\rm cm}$ and $\phi$
shows that absolute values of radiative corrections do not exceed
1.5 \%. The procedure of sampling the  radiative corrections with analytical
functions has been developed allowing for flexible and rapid 
applications of the corrections in the data analysis. 
\end{abstract}
%
%
\section{Introduction}
Evaluation of radiative corrections (RC) to the observables of
the exclusive pion electroproduction
\begin{equation}
\label{react1}
e^- + p \to e^- + p + \pi^{\circ}
\end{equation}
is, as a rule, performed in the framework of the approach
developed in Ref.~\cite{afana02}. Practical calculations, which
involve application of the code {\small EXCLURAD}~\cite{exclu},
take a noticeable amount of time even in the simple case of several
kinematic points. As a result, the analysis of the data registered
in a wide range of kinematic variables may be hindered considerably.
 Approximations of the results obtained from {\small EXCLURAD}
with analytical functions can substantially decrease time
of the data analysis and maintain a high level of accuracy in
the applied corrections. An input for the approximations is obtained
by sampling of the energy and angular dependence of RC and the 
single spin asymmetries $A$ and $A_{\rm RC}$
evaluated in the Born approximation and by including all relevant
radiative processes, respectively.
%
\section{Kinematics}
\begin{figure}[t]
\vspace*{-.8cm}
  \hfil
   \begin{minipage}[t]{0.5 \linewidth}
\begin{center}
\hspace*{-0.1cm}\mbox{\epsfysize=\hsize\epsffile{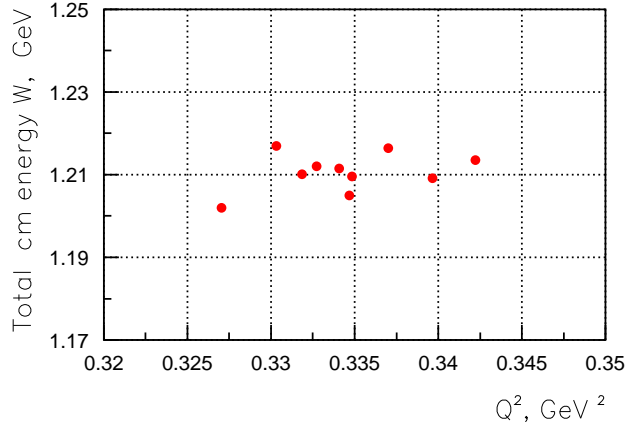}}
\end{center}
  \hfil
   \end{minipage}
\vspace{-.8cm}
\caption{Location of the mean values of $W$ and $Q^2$ corresponding
to different experimental settings in the measurements
of the single spin asymmetry at MAMI. 
\label{fig:kinema}}
\end{figure}

The data on single spin asymmetry, which is observed in the
reaction~(\ref{react1}) in polarized electron beam 
has been collected at MAMI microtron of Mainz~\cite{mainz03}.
The relevant kinematic region is  shown in Fig.~\ref{fig:kinema} 
as a scatter plot with ten points on the $W - Q^2$ plane. 
Position of the points is determined by the incident
beam energy $E$ = 883 MeV, and true average of each experimental
 bin has been found in the analysis of Ref.~\cite{october03}.
%

The code {\small EXCLURAD} computes RC to the fourfold
cross section $d^4/dWdQ^2 d{\rm cos}\theta_h d \phi_{h}$ and to
polarization beam asymmetry for the reaction~(\ref{react1}).
The cm system angles $\theta_h$ and $\phi_{h}$ are between
the direction of the virtual photon momentum 
($\vec q  = \vec k_1 - \vec k_2$) and momentum of the final pion.
As long as a pion in reaction~(\ref{react1}) is the undetected
hadron, the angles $\theta_h$ and $\phi_{h}$ are defined with
respect to the direction of $- \vec p~'$, i.e. {\em opposite}
to the recoil proton momentum~\cite{afana02}. In what follows these 
angles are denoted as $\theta_{\rm cm}$ and $\phi$.

Calculations of RC, $A$ and $A_{\rm RC}$ have been performed 
at the fixed value of the
virtual photon transfered momentum --- $Q^2$ = 0.335 GeV$^2$, which
corresponded to the central value of the $Q^2$ region represented
by the scatter plot in Fig.~\ref{fig:kinema}.
With the adopted definition of $\phi$, 
both  $A(\phi)$ and $A_{\rm RC}(\phi)$ are symmetric
with respect to the angle $\phi = 180^{\circ}$.
In the experiment,  $\theta_{\rm cm}$~ covered the range 
$2.5^{\circ} \leq \theta_{\rm cm} \leq 37.5^{\circ}$, and the asymmetries
were sampled in 10 points inside this interval.
The sampling of the $\phi$ dependence has been done in 11 points inside
in the interval from 0 to 180$^{\circ}$.

As has been demonstrated in Ref.~\cite{afana02}, the inclusive asymmetry  
for the reaction~(\ref{react1})
has a nontrivial dependence on the center of mass energy $W$
around $W \approx 1.3$~GeV where it changes sign. 
Therefore, for reliable evaluation of RC one has to sample 
a much wider range of $W$ than that covered by average values in 
Fig.~\ref{fig:kinema}. The relevant values corresponded to
 $W$ = 1.10, 1.15, 1.21, 1.25 and 1.30 GeV.

%
%
\section{Analytical expressions approximating evaluated RC}

\subsection{$W$ dependence}

The Bezier curves are used as the basis for approximations of
both Born asymmetry $A_0(\phi,\theta,W)$
and the asymmetry modified by the radiative effects --- 
$A_{\rm RC}(\phi,\theta,W)$. They proved to yield a simple 
approximating tool in the entire range of the kinematics under
consideration.
In this approach the asymmetry evaluated by the {\small EXCLURAD} 
code in a number of fixed kinematic points is approximated as
\begin{equation}
\label{b-sum}
A = \sum_{l=0}^3 P_l B_l(u)~,
\end{equation}
where $ P_l$ are parameters of the fit and $B_l(u)$ are functions 
defined on the interval  $\{0, 1\}$:
\begin{eqnarray}
\label{bezier}
B_0 &= &(1-u)^3 ,\nonumber\\
B_1 &= &3u(1-u)^2 ,\nonumber\\
B_2 &= &3u^2(1-u) ,\\
B_3 &= &u^3 .\nonumber\\
\nonumber
\end{eqnarray}
Due to this particular choice of functions $B_l$, $P_0$ and $P_3$ coincide
with the values of $A$ on the left and right boundary of the
interval, respectively. Thus, it is only $P_1$ and $P_2$, which are
to be determined from the fit.

When fitting $W$ dependence of the asymmetry evaluated on the interval
\{$W_{\rm min}, W_{\rm max}$\},  the following change was done:
$$
u ={ W - W_{\rm min} \over  W_{\rm max} -  W_{\rm min} }~. 
$$
For the applications in the Mainz experiment, $u$ has been 
calculated by taking $W_{\rm min}$ = 1.10 GeV and 
$W_{\rm max}$ = 1.30 GeV. 

\subsection{Dependence on the azimuthal angle $\phi$}
Similarly to the case of $W$ dependence, one has to change
$u =\phi / 180^\circ $  in Eq.~(\ref{b-sum}) in the fitting
of $A(\phi)$. 
On the other hand, one can suggest some simplifications to Eq.~(\ref{b-sum}),
which would better suit conditions of the Mainz experiment, in which 
the difference between Born and RC asymmetries is very small
everywhere except of $W \sim 1.3$~GeV. This is done by noticing that
 $A(\phi)$ is close to a sine function and by replacing 1st and 4th
terms (which have to disappear in the case of purely sine function)
with a sine term:
\begin{equation}
\label{sum-3}
A(\phi)  =  P_0 {\rm sin}(\phi) +   P_1 B_1(u) + P_2 B_2(u),
\end{equation}
where  $u =\phi / 180^\circ $.
The dependence of parameters $P_l$ on $W$ is obtained by fitting
the results with Eq.~(\ref{b-sum}). The resulting 2D asymmetry is
given by
\begin{equation}
\label{phi-w}
A(\phi,W)  =  \sum_{l=0}^2 B_l^{\phi} \sum_{m=0}^3 P_{lm} B_m^W~,  
\end{equation}
where $B_0^{\phi}$ = sin$(\phi)$, and $ B_{1,2}^{\phi}$ are usual functions
defined in Eq.~(\ref{bezier}).

\subsection{Dependence on the angle $\theta_{\rm cm}$}

As it is obtained from calculations of the  $A(\theta_{\rm cm})$ by
the {\small EXCLURAD}, both  $A(\theta_{\rm cm})$ and 
$A_{\rm RC}(\theta_{\rm cm})$ are small
and slowly varying functions in the kinematics covered by the 
experiment ($2.5^{\circ} \leq \theta_{\rm cm} \leq 37.5^{\circ}$).
Therefore, the most economical way of approximating the asymmetry
is the choice of the two-terms functions as follows:
\begin{equation}
\label{sum-2}
A(\theta) = a {\rm sin}(\theta_{\rm cm}) + b {\rm sin}^2(\theta_{\rm cm})~.
\end{equation}
Correspondingly, approximation of the $A(\phi,\theta_{\rm cm})$
has been performed for a given value of $W$ in the
following form:
\begin{equation}
\label{phi-th}
A(\phi,\theta_{\rm cm}) = \sum_{l=0}^1 
({\rm sin}(\theta_{\rm cm}))^{l+1} \sum_{m=0}^3 P_{lm} B_m^{\phi}~,  
\end{equation}
where functions $B_m^{\phi}$ are defined in Eq.~(\ref{bezier}). 
Parameters $P_{lm}$ of the fit performed for the case 
with radiative processes are shown in Table 1.

\begin{table}[h]
\centerline{\begin{tabular}{|l|c|c|c|c|}
\hline
 & m = 0 &  m = 1 &  m = 2 &  m = 3 \\ 
\hline
l = 0 & --0.51822 & 8.2071 & 6.8661 & --0.51799 \\
l = 1 &  1.3751 & --4.0265 & --6.4145 &  1.3751 \\
\hline
\end{tabular}}
\caption{Parameters $P_{lm}$ that determine 
$A_{\rm RC}(\phi,\theta_{\rm cm})$ }
\end{table}

%
\begin{figure}[ht]
  \hfil
   \begin{minipage}[tbh]{0.55 \linewidth}
\begin{center}
\hspace*{-0.1cm}\mbox{\epsfysize=\hsize\epsffile{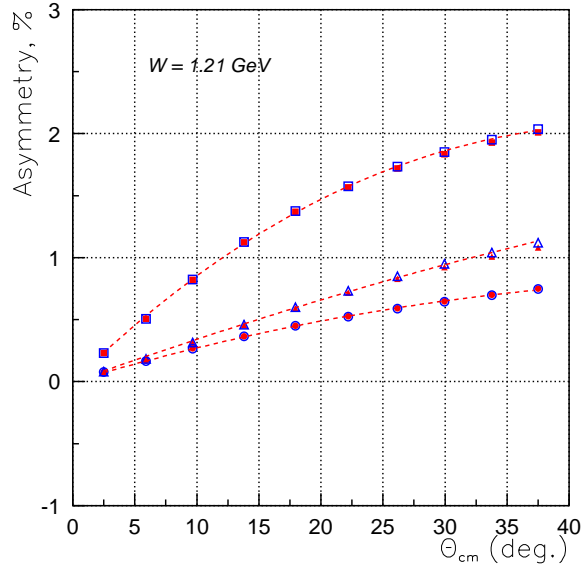}}
\end{center}
  \hfil
   \end{minipage}
\vspace{-0.8cm}
\caption{Asymmetry as a function of $\theta_{\rm cm}$
 evaluated for three values of the  azimuthal
angle $\phi$ --- 20$^{\circ}$ (triangles),  90$^{\circ}$ (squares)
and 160$^{\circ}$ (circles). Asymmetries calculated in Born 
approximation and by accounting for radiative precesses 
(assuming $\Delta S = 0.02$ GeV$^2$) are
shown with filled and open symbols, respectively. 
\label{fig:rc-theta}}
\end{figure}
%
%
\begin{figure}[tbh]
\hspace*{3.8cm}\mbox{\epsfxsize=.55\hsize\epsffile{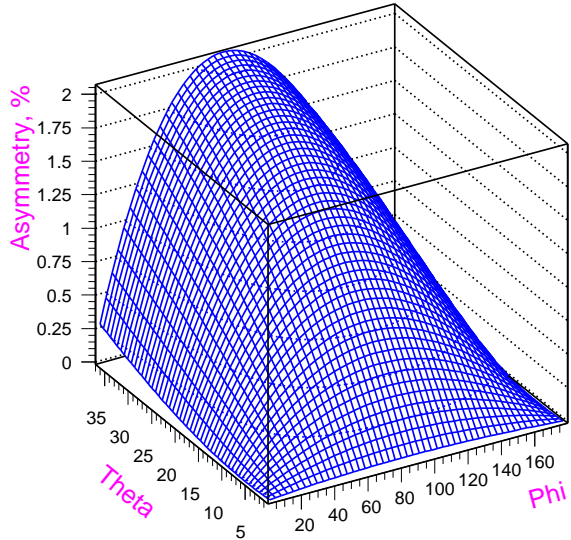}}
\vspace*{-1cm}
\caption{Asymmetry  $A_{\rm RC}(\phi,\theta_{\rm cm})$ evaluated at
$W$ = 1.21 GeV by assuming $\Delta S =  0.02$ GeV$^2$. 
\label{fig:2d-phi-th}}
\end{figure}


\section{Results}
It was found that the modification of the asymmetry due to
radiative effects in the reaction under study is the strongest at 
$\theta_{\rm cm}$ = 37.5$^{\circ}$. This is demonstrated by the
results of calculations $A(\theta_{\rm cm})$ and 
$A_{\rm RC}(\theta_{\rm cm})$ shown in  Fig.~\ref{fig:rc-theta}
for three values of the azimuthal angle~$\phi$.
%

Radiative corrections depend on the mass window $\Delta S$ around
the missing mass peak used in the event selection procedure if 
$\Delta S < \nu_{\rm max}$, where $\nu_{\rm max}$ is introduced 
in {\small EXCLURAD} as the  upper limit of integration in
calculating the cross section of the radiative process.
The parameter $\nu_{\rm max}$ describes the missing mass of the
radiative origin (due to undetected~$\gamma$). It is dubbed as 
``maximum inelasticity'' and is defined in Ref.~\cite{afana02} as
\begin{eqnarray}
\nu_{\rm max} &&= (W-M_p)^2 - m_{\pi^{\circ}}^2~. 
\end{eqnarray}
Undetected gammas
additionally distort the missing mass spectrum, which is
of the hadronic origin (undetected $\pi^0$ in the considered case).
 The selection cuts, which are usually applied to the distribution of the
events versus missing mass squared $M_x^2$, are generally more tight
than $\Delta S = \nu_{\rm max}$, which is modeled in {\small EXCLURAD}
by replacing the  upper limit of integration with  $\nu_{\rm cut}$.
Accordingly, the mass window $\Delta S$ is related to $\nu_{\rm cut}$
as
\begin{eqnarray}
\Delta S &&= \nu_{\rm max} - \nu_{\rm cut}~.\\ \nonumber
\end{eqnarray}
The corrections
naturally increase as the cuts tighten, corresponding to 
narrowing of the  mass window $\Delta S$.

\begin{figure}[h]
\vspace{-0.3cm}
 \hfil
   \begin{minipage}[t]{0.5 \linewidth}
\begin{center}
\hspace*{-0.1cm}\mbox{\epsfysize=\hsize\epsffile{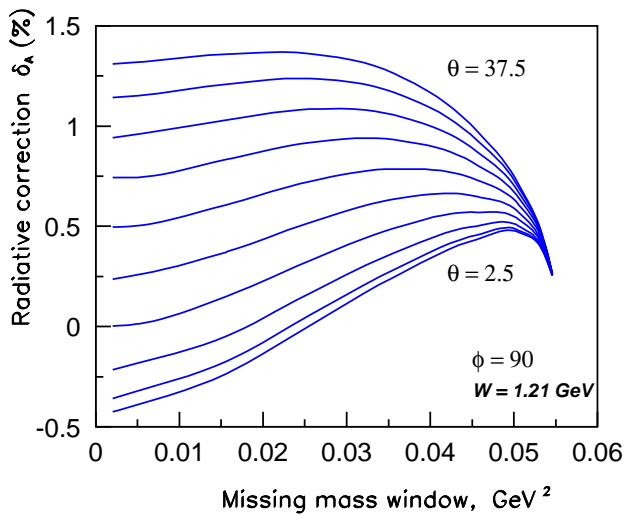}}
\end{center}
   \end{minipage}
\vspace{-0.3cm}
\caption{Radiative corrections as a function of the $\Delta S$
window evaluated in 10  points of the phase space corresponding to the 
angle $\theta_{\rm cm}$ in the range 2.5$^\circ$ -- 37.5$^\circ$
and at the azimuthal angle $\phi$ = 90$^\circ$.
\label{fig:rc-theta-wind}}
\end{figure}

The two-dimensional plot for $A_{\rm RC}(\theta_{\rm cm})$
corresponding to the case of $\Delta S  = 0.02$  GeV$^2$ 
is obtained analytically by employing
Eq.~(\ref{phi-th}). It is displayed in Fig.~\ref{fig:2d-phi-th}.
The difference between  $A_{\rm RC}(\theta_{\rm cm})$ and 
$A(\theta_{\rm cm})$ is better seen from 
Fig.~\ref{fig:rc-theta-wind}, which shows radiative 
corrections $\delta_A$ defined in Ref.~\cite{afana02} as 
\begin{equation}
\label{radcor}
\delta_A = { A_{\rm RC} - A \over A}\cdot 100\%~.
\end{equation}
The set of ten curves  in Fig.~\ref{fig:rc-theta-wind}
shows the dependence of $\delta_A$ 
for ten values of $\theta_{\rm cm}$ on the
size of the $\Delta S$ window used in the event selection procedure.
The results have been produced for the azimuthal angle 
$\phi$ = 90$^\circ$, where corrections are found to be largest.

%
%
In this study,
a larger number of results have been produced for the 
kinematics corresponding to $\theta_{\rm cm}$ = 37.5$^{\circ}$.

The curves in Fig.~5 represent
results of the fits, in which Eq.~(\ref{phi-w}) was employed.
The fitting procedure started from  Eq.~(\ref{sum-3}) and
 three parameters $P_l$ have been obtained in five
points of $W$, which allowed to conduct a fit of the $W$ dependence
for each of them. It is realized by using Eq.~(\ref{b-sum}), and
 Fig.~\ref{wdfitborn} shows obtained parameters.

\begin{table}
\centerline{\begin{tabular}{|l|c|c|c|c|}
\hline
 & m = 0 &  m = 1 &  m = 2 &  m = 3 \\ 
\hline
l = 0 & --1.4506 & --4.6186 & --1.9493 & 0.17369 \\
l = 1 & 5.7645 & 12.049  & 7.0872  & --2.7053 \\
l = 2 & 2.4684 & 6.8852  & 8.1888 &  --2.6894\\ 
\hline
\end{tabular}}
\caption{Parameters $P_{lm}$ that determine $A(\phi,W)$ }
\end{table}
\begin{table}
\centerline{\begin{tabular}{|l|c|c|c|c|}
\hline
 & m = 0 &  m = 1 &  m = 2 &  m = 3 \\ 
\hline
l = 0 & --1.4544 & --4.6810 & --2.2870 & --0.54204 \\
l = 1 & 5.7668 & 12.365 & 6.2068 & 0.45351 \\
l = 2 & 2.4694 & 7.1451 & 7.6380 & --0.64024 \\ 
\hline
\end{tabular}}
\caption{Parameters $P_{lm}$ that determine $A_{\rm RC}(\phi,W)$ }
\end{table}

The parameters thus obtained are presented in Tables~2 and~3. They are
used in Eq.~(\ref{phi-w}) for rapid evaluation of $A(\phi, W)$
and   $A_{\rm RC}(\phi, W)$ at
the angle $\theta_{\rm cm}$~= 37.5$^\circ$ and the $\Delta S$ window of
0.020 GeV$^2$. The result is displayed in Fig.~\ref{born2d}.
\begin{table}
\centerline{\begin{tabular}{|l|c|c|c|c|}
\hline
 & m = 0 &  m = 1 &  m = 2 &  m = 3 \\ 
\hline
l = 0 & --1.4544 & --4.6810 & --2.2870 & --0.54204 \\
l = 1 & 5.7668 & 12.365 & 6.2068 & 0.45351 \\
l = 2 & 2.4694 & 7.1451 & 7.6380 & --0.64024 \\ 
\hline
\end{tabular}}
\caption{Parameters $P_{lm}$ that determine $A_{\rm RC}(\phi,W)$ }
\end{table}

\setcounter{figure}{5}
\begin{center}
\begin{figure}[p]
\begin{center}
\begin{minipage}[t]{0.5 \linewidth}
\hspace*{-3cm}\mbox{\epsfysize=\hsize\epsffile{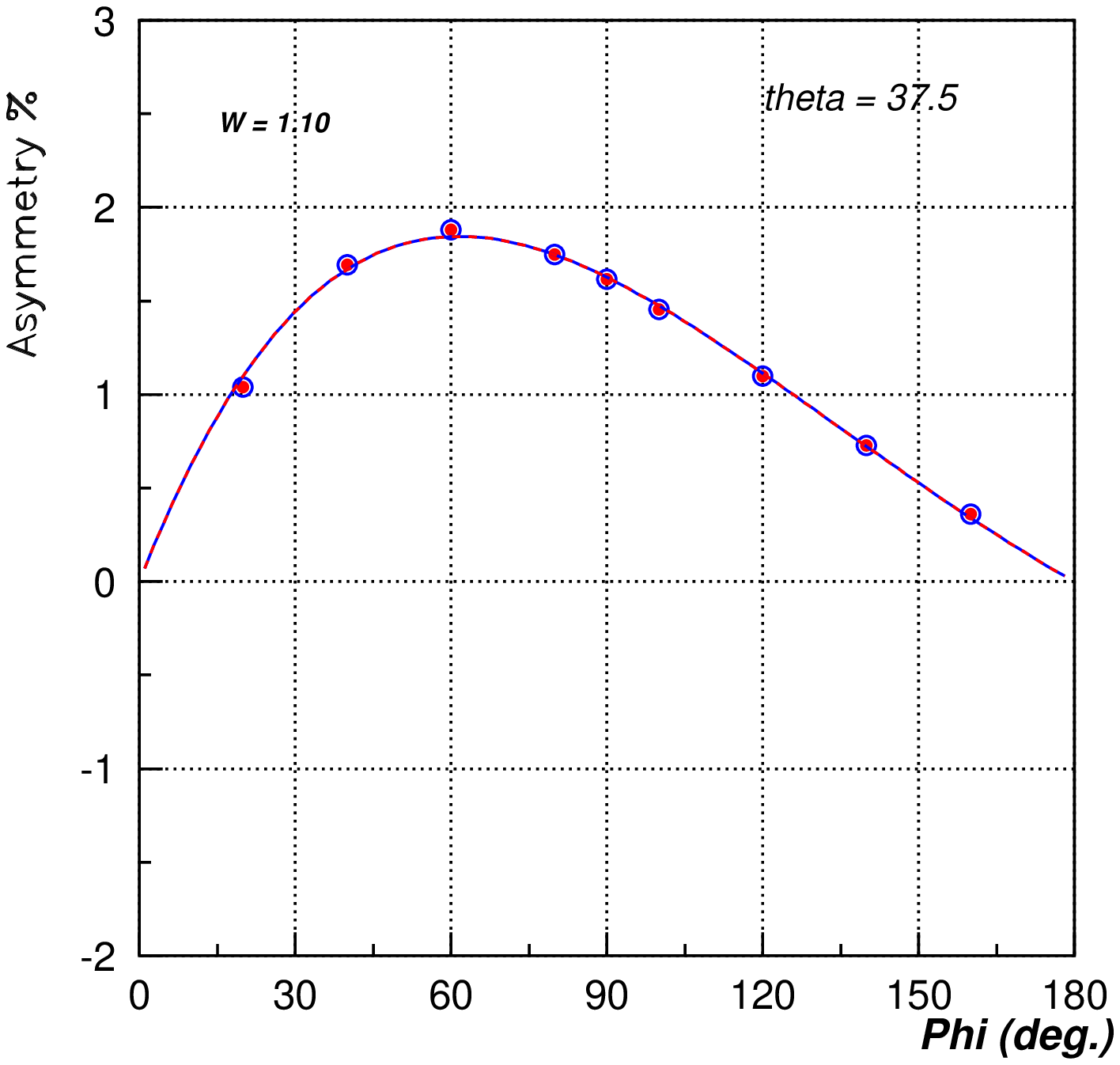}
\hspace{-0.5cm}\epsfysize=
\hsize\epsffile{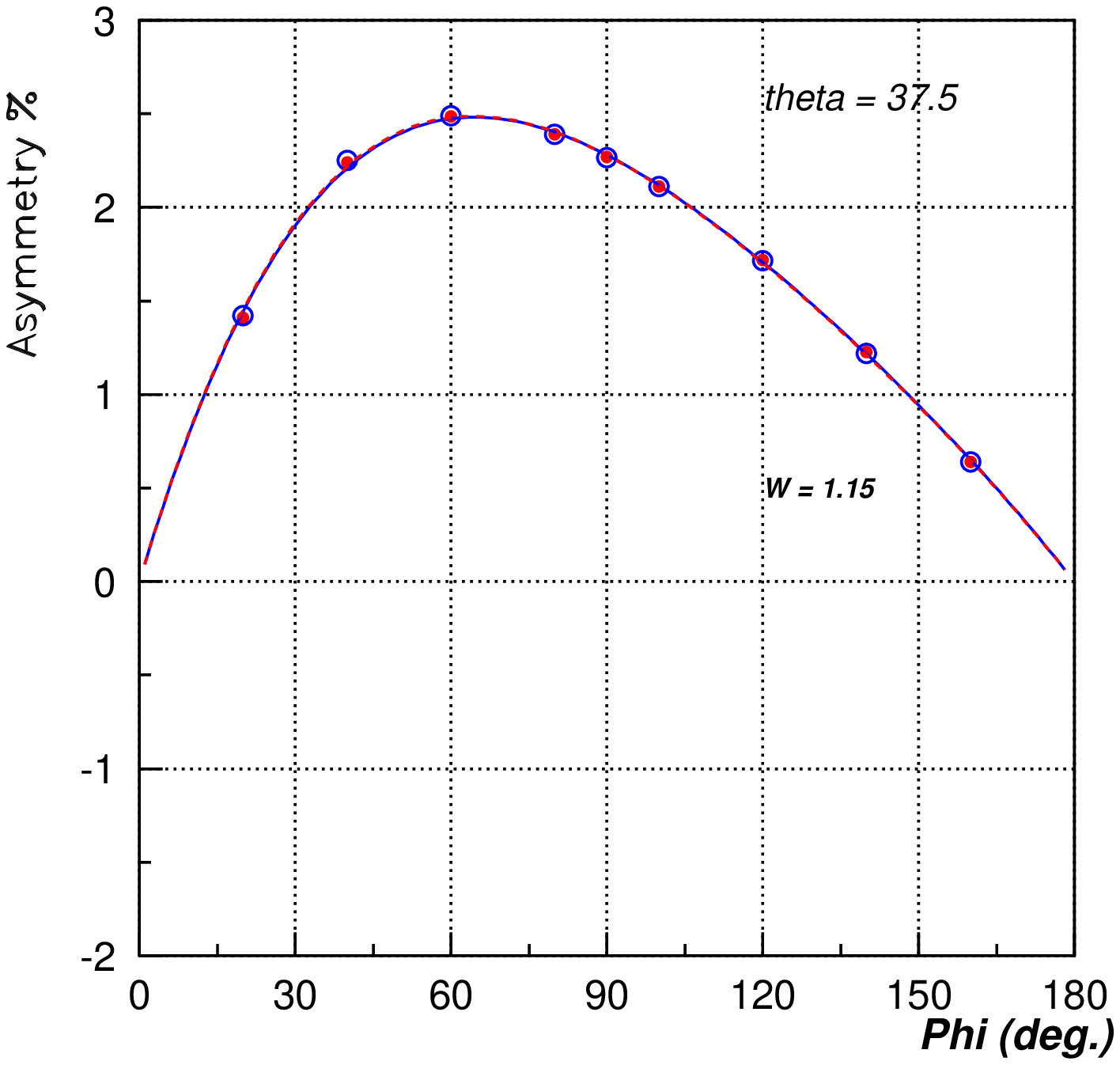}}
\end{minipage}
\end{center}
\vspace{-1.0cm}
\begin{center}
\begin{minipage}[t]{0.5   \linewidth}
\hspace*{-3cm}\mbox{\epsfysize=\hsize\epsffile{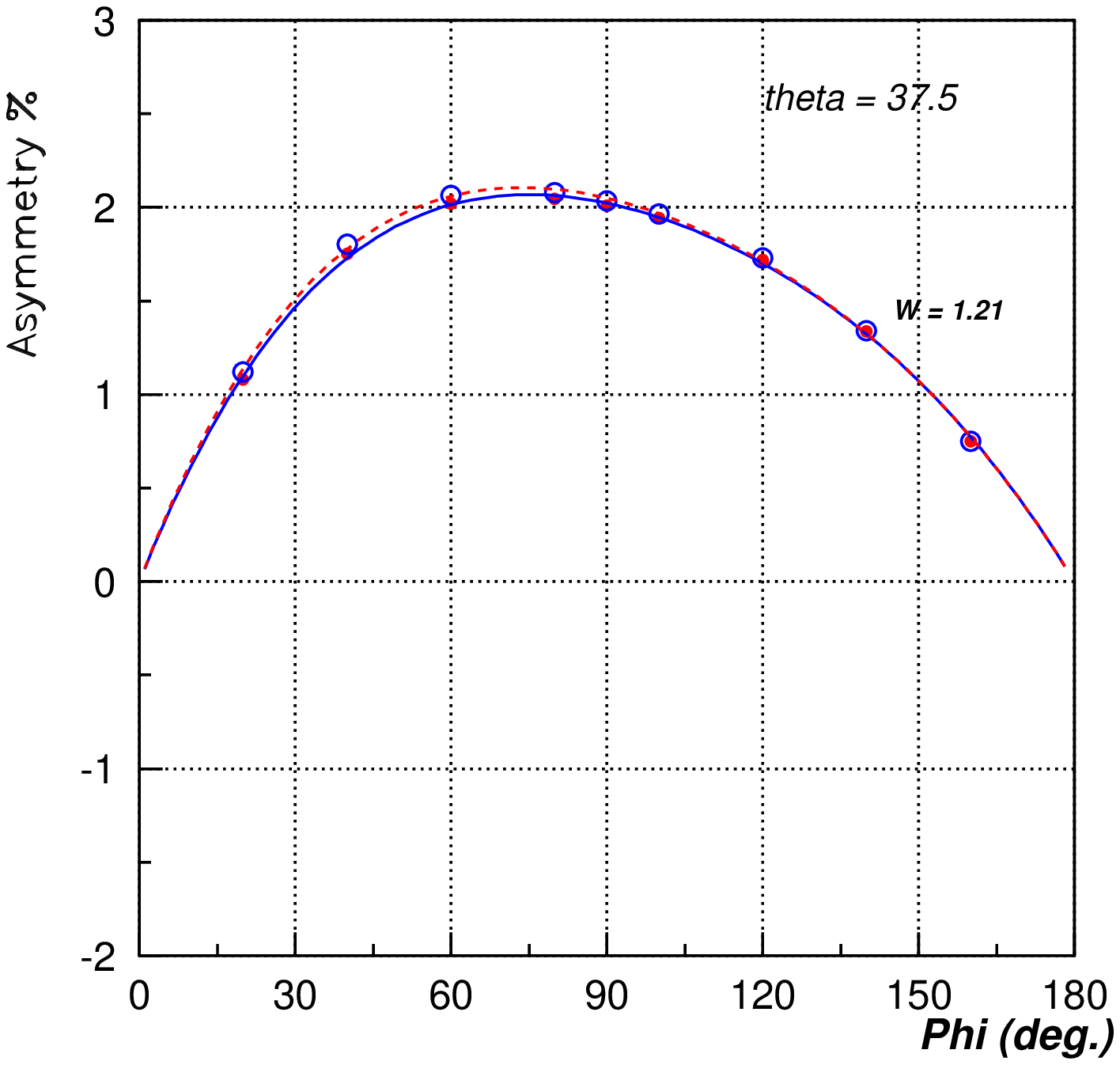}
\hspace{-0.5cm}\epsfysize=
\hsize\epsffile{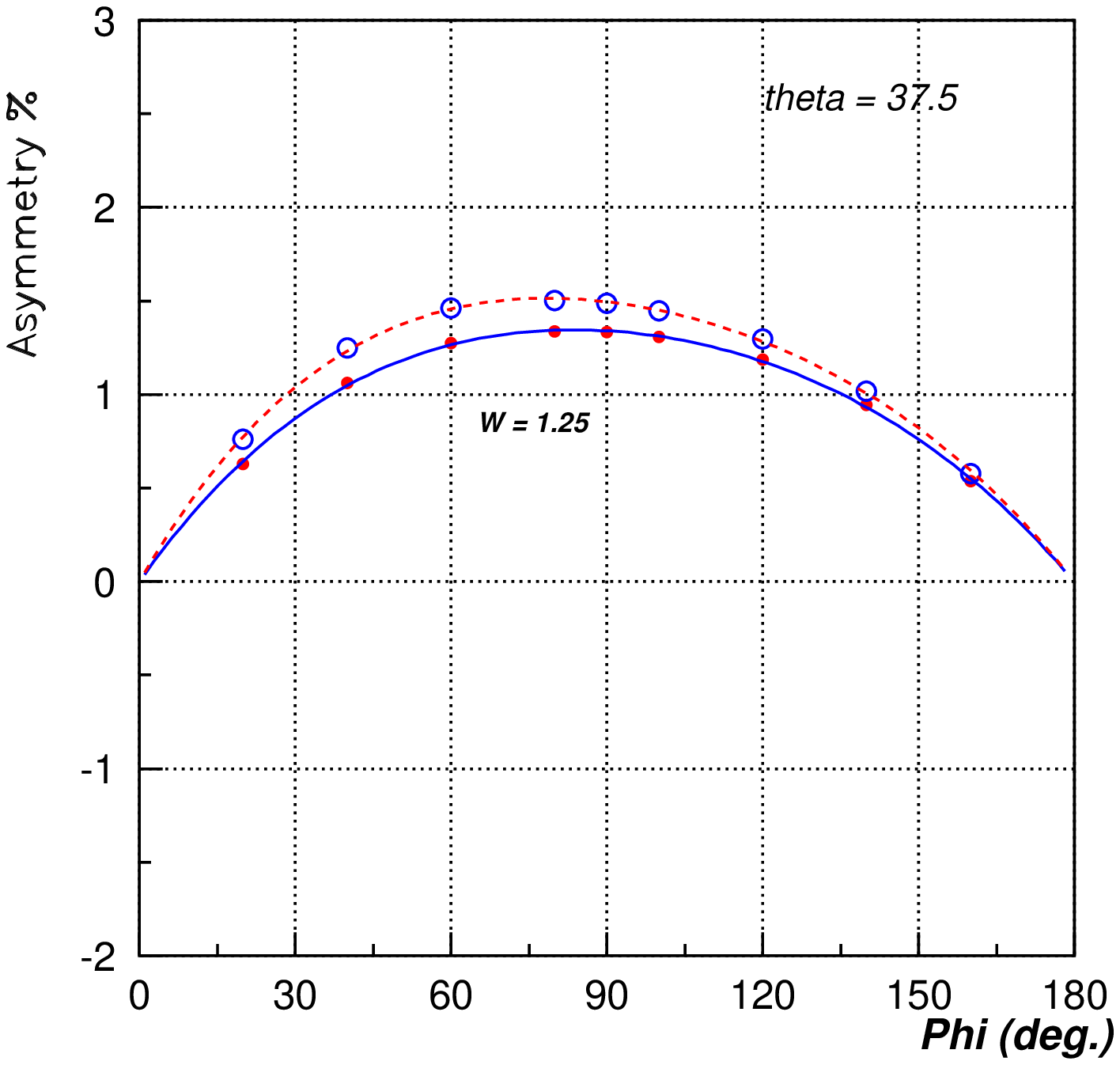}}
\end{minipage}
\end{center}
\vspace{-1.0cm}
\begin{center}
\begin{minipage}[t]{0.5   \linewidth}
\hspace*{-3cm}\mbox{\epsfysize=\hsize\epsffile{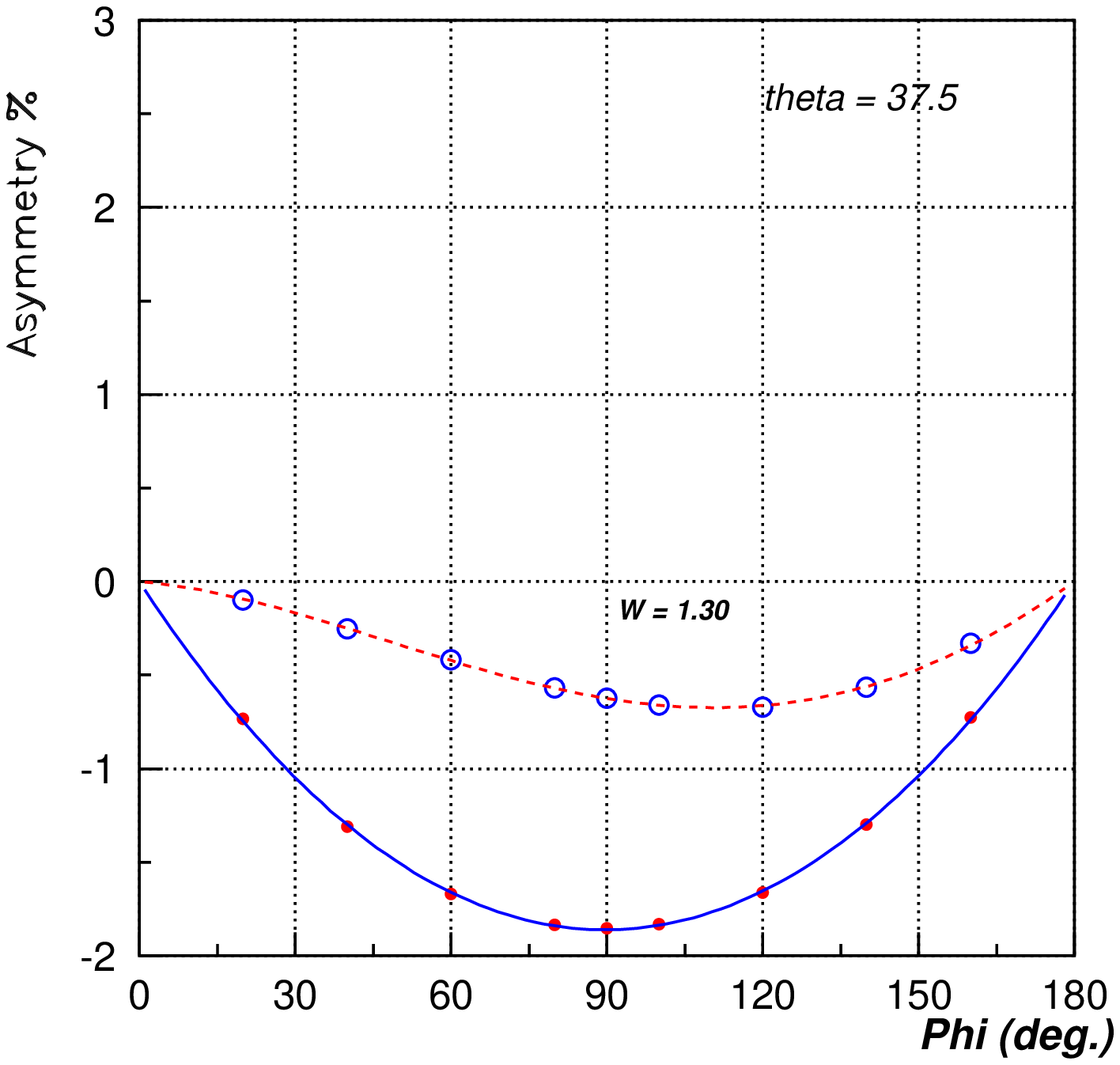}
\hspace{0.5cm}\parbox{6.5cm}{\vspace*{-8.0cm}Fig.~5 Born-term -- $A$, and
RC-smeared -- $A_{\rm RC}$, asymmetries used for the evaluation of radiative
corrections $\delta_A$
as a function of the azimuthal angle $\phi$ at five values
of the total c.m. energy $W$ -- 1.10, 1.15, 1.21 1.25 and 1.30 GeV.
~In addition, $\theta_{\rm cm}$ = 37.5$^{\circ}$ and the $\Delta S$
window is 0.020 GeV$^2$. Full and open points correspond to exact
calculations of the  asymmetries $A$ and $A_{\rm RC}$, respectively.
They are approximated by full and dashed lines.}}
\end{minipage}
\end{center}
\end{figure}
\end{center}
%
%
%
Dependence of the additive correction defined as $A - A_{\rm RC}$ on 
$\phi$ and $W$ is shown in Fig.~\ref{diff} for the same
conditions as in Fig.~\ref{born2d}.
%
%
\begin{figure}[tbh]
\hspace*{4.5cm}\mbox{\epsfxsize=.45\hsize\epsffile{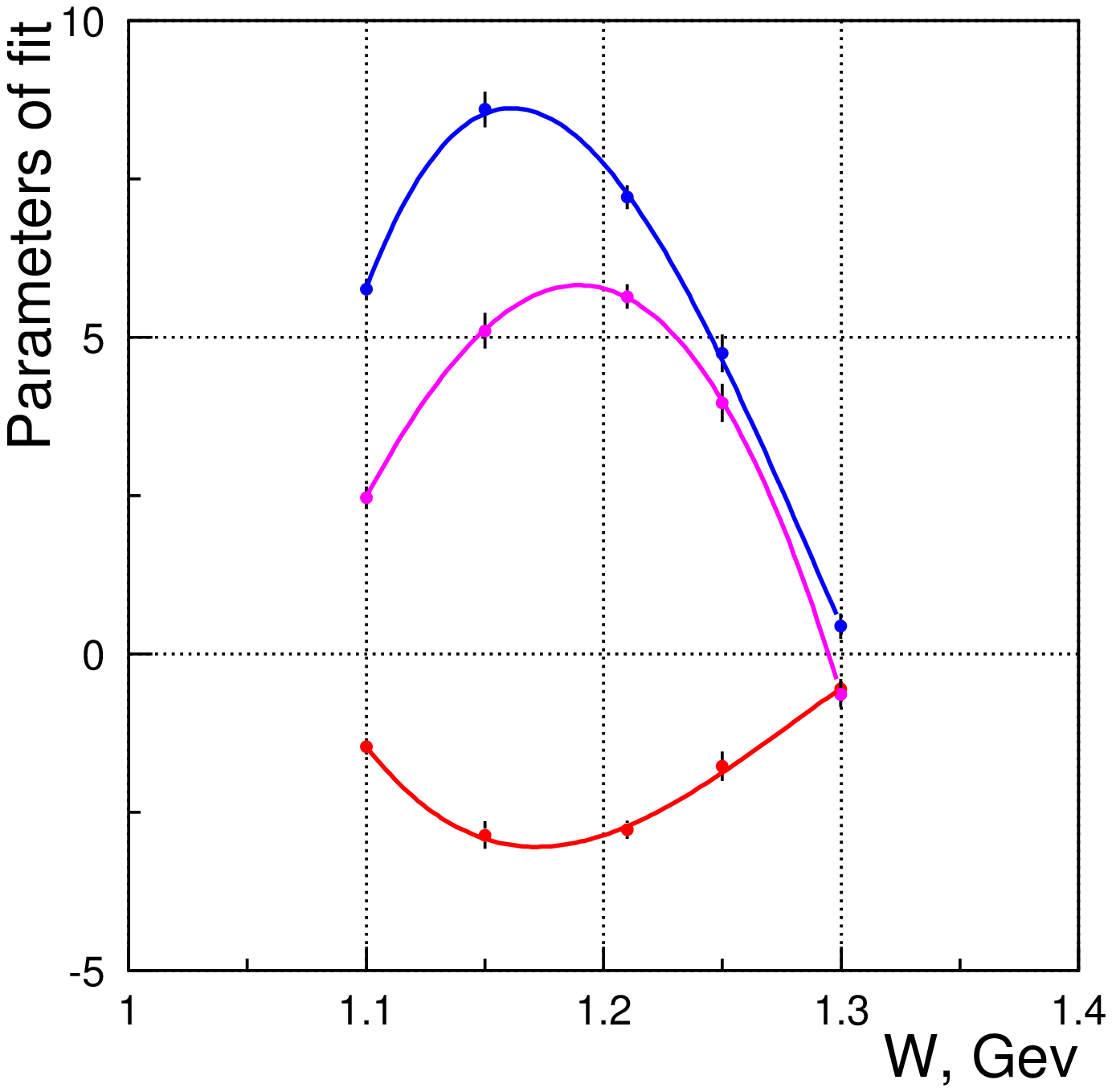}}
\vspace{-0.3cm}
\caption{$W$ dependence of parameters $P_l$ of Eq.~(3) evaluated 
for Born asymmetry at $\theta_{\rm cm}$~= 37.5$^\circ$. Three curves
(from top to bottom) correspond to $P_1$,  $P_2$ and  $P_0$. 
\label{wdfitborn}} 
\end{figure}
As it is demonstrated by Fig.~\ref{fig:rc-theta}, the asymmetries 
$ A(\phi,\theta_{\rm cm})$ and $A_{\rm RC}(\phi,\theta_{\rm cm})$
are slow varying functions  of $\theta_{\rm cm}$, which reach maximum 
at the end of the considered kinematic interval -- 
$\theta_{\rm cm}$ = 37.5$^{\circ}$.
This feature justifies  an approximate sampling of asymmetries 
for any given value of $\theta_{\rm cm}$ by scaling down the
asymmetry found for $\theta_{\rm cm}$ = 37.5$^{\circ}$.
To this end,
one can construct a normalizing function $f_N(\phi,\theta_{cm})$
\begin{equation}
\label{2d_norm}
f_N(\phi,\theta_{\rm cm}) = {  A(\phi,\theta_{\rm cm}) \over  
 A(\phi,\theta_{\rm cm}^{\rm max})} ~,
\end{equation}
where  $A(\phi,\theta_{\rm cm})$ is defined by Eq.~(\ref{phi-th}), and 
$ A(\phi,\theta_{\rm cm}^{\rm max})$ corresponds to the results of the 
fit at $\theta_{\rm cm}$ = 37.5$^{\circ}$.
The combination of  $f_N(\phi,\theta_{\rm cm})$ with the results given
by Eq.~(\ref{phi-w}) provides a simple and precise tool for evaluating 
$A(\phi, W, \theta_{\rm cm})$ in the region around $W$ = 1.2 GeV:
\begin{equation}
\label{3d_asym}
A(\phi, W, \theta_{\rm cm}) = A(\phi, W) f_N(\phi,\theta_{\rm cm}). 
\end{equation}
\vfill
\eject

%
%
%
\begin{figure}[tbh]
 \vspace*{-1cm}
  \hfil
   \begin{minipage}[h]{0.45\linewidth}
\begin{center}
\hspace*{-3.5cm}\mbox{\epsfysize=\hsize\epsffile{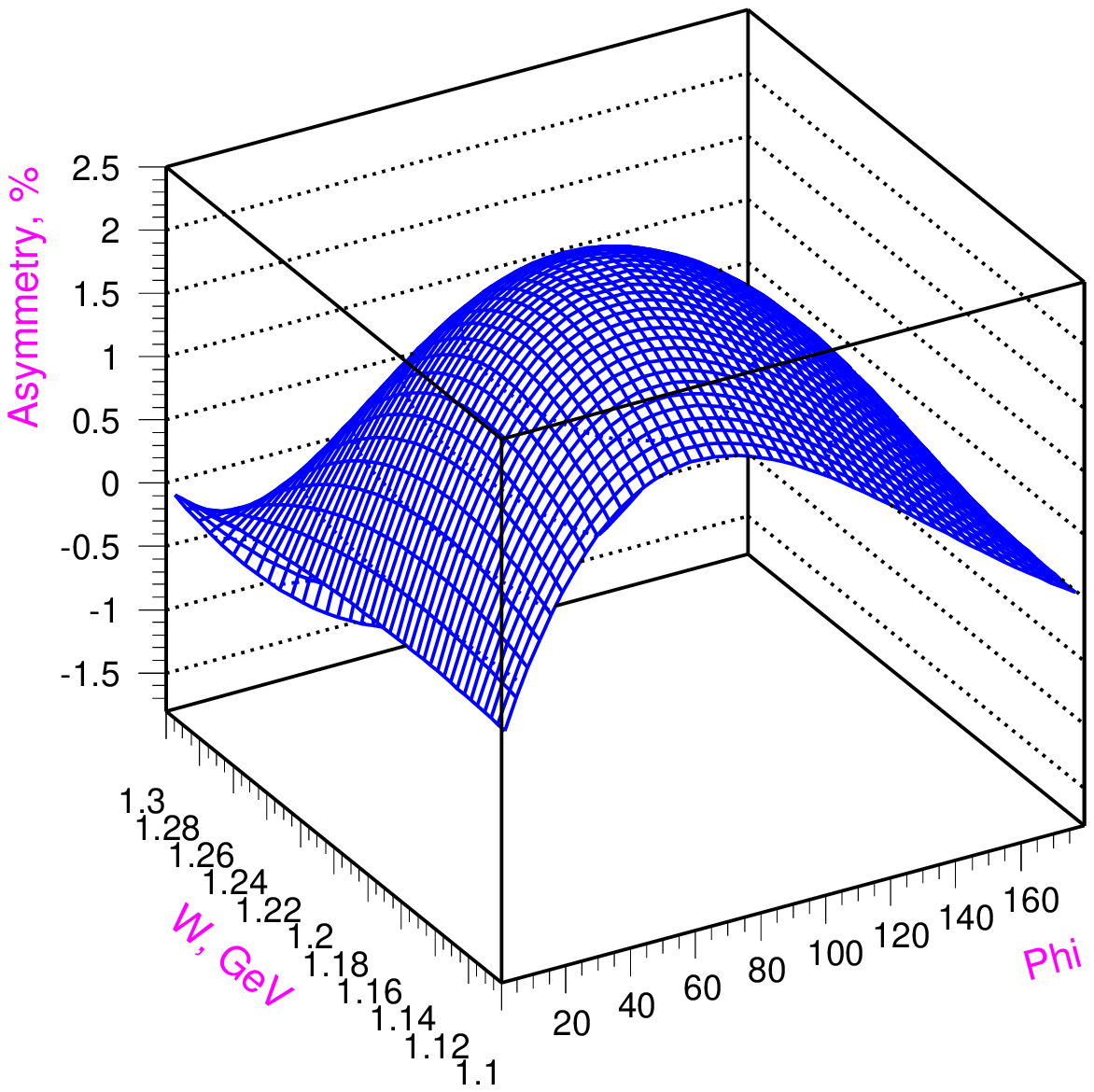}
\hspace*{-0.1cm}\epsfysize=\hsize\epsffile{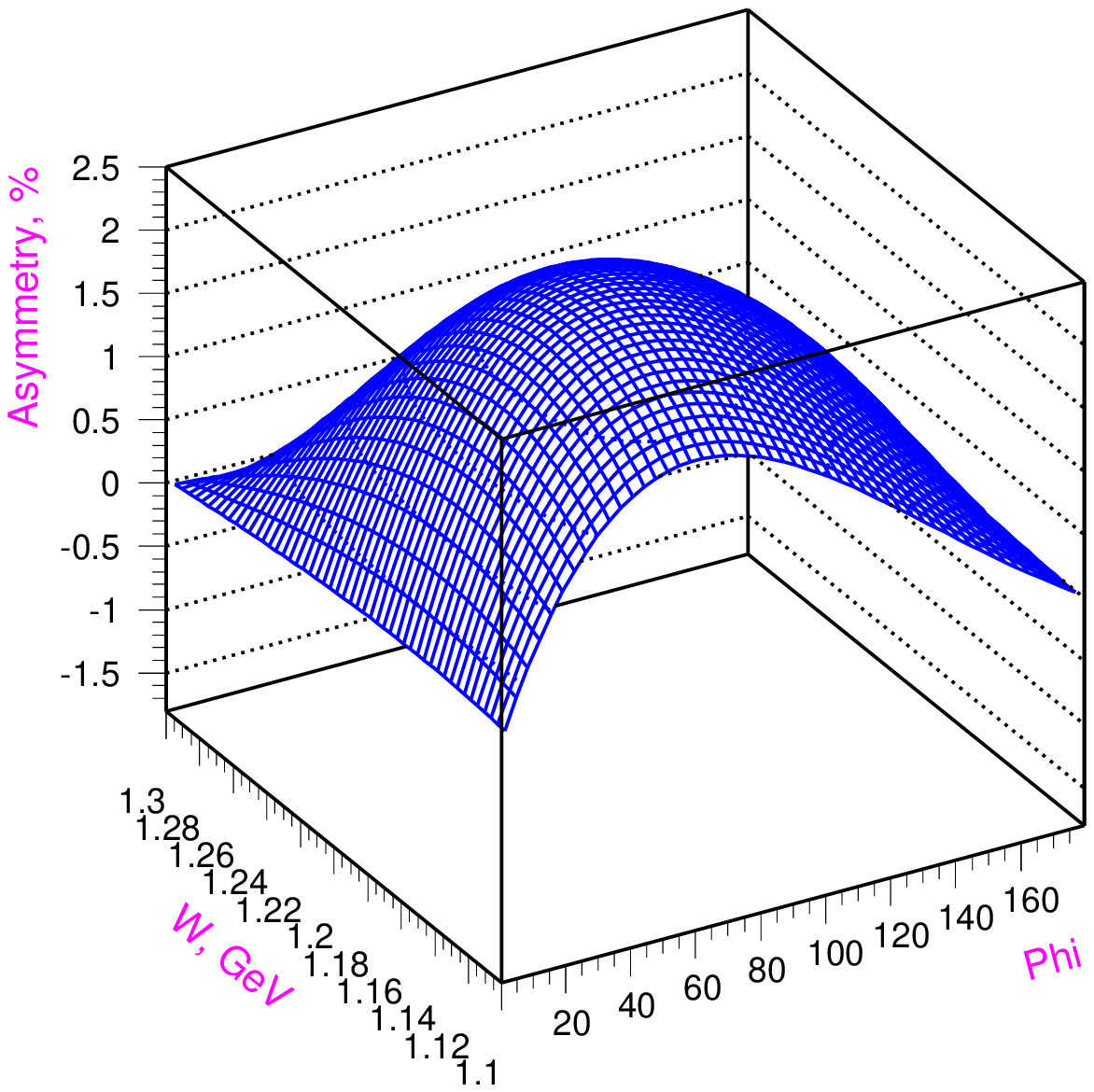} }
\end{center}
  \hfil
   \end{minipage}
\vspace{-1cm}
\caption{Single spin Born asymmetry $A(\phi,W)$ (left panel) and the 
asymmetry $A_{\rm RC}(\phi,W)$ modified by radiative effects (right panel),
both evaluated for $\theta_{\rm cm}$~= 37.5$^\circ$.
\label{born2d}}
\end{figure}
%
%
\begin{figure}[hb]
 \vspace*{-0.5cm}
\hspace*{3.5cm}\mbox{\epsfxsize=.55\hsize\epsffile{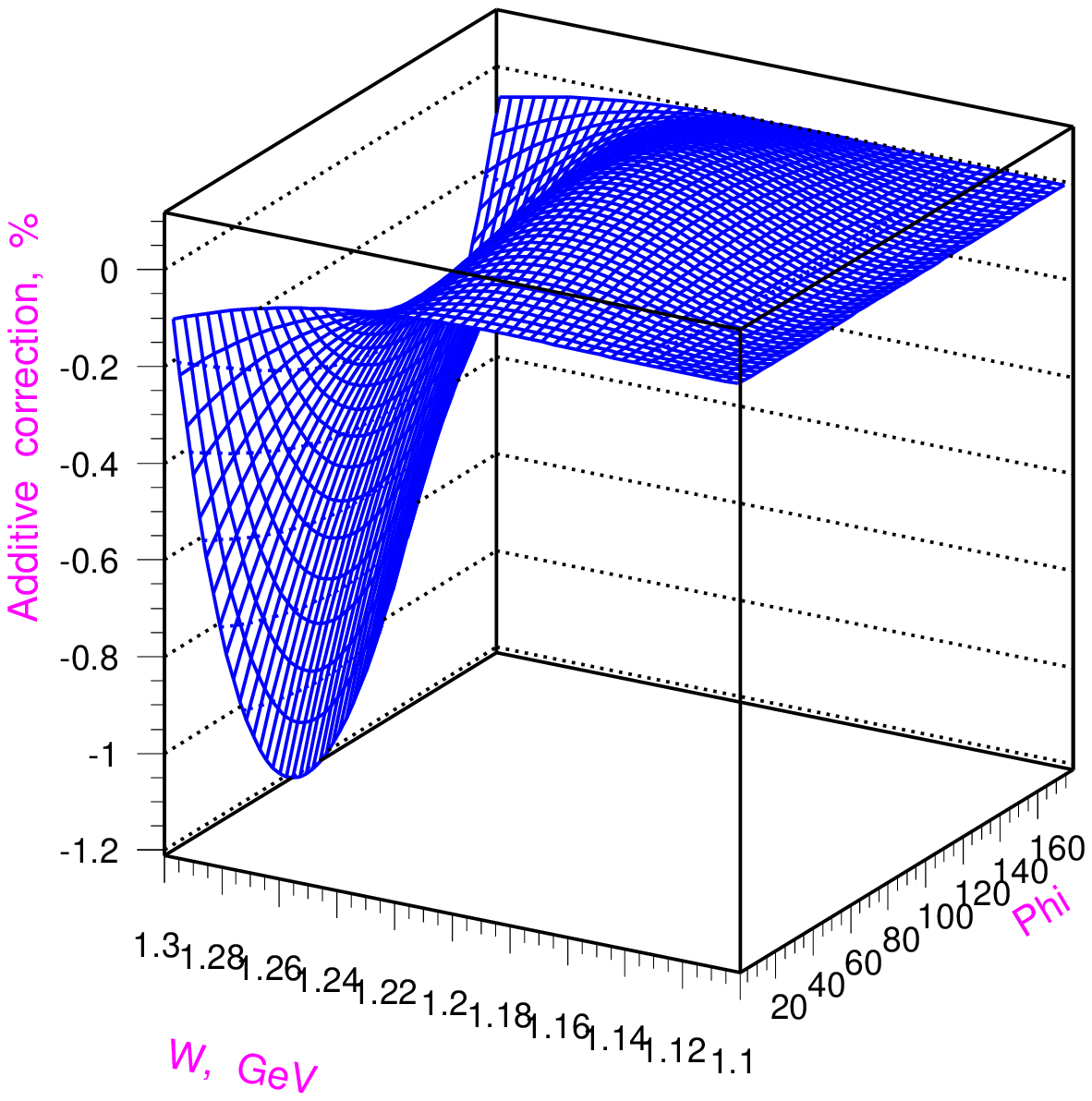}} 
\vspace{-0.5cm}
\caption{ Additive correction obtained as $A - A_{\rm RC}$, which
corresponds to the difference between the plots in the left and
right panels of Fig.~7.
\label{diff}}
\end{figure}
%

\begin{center}
\begin{figure}[tbh]
  \hfil
   \begin{minipage}[h]{1.03 \linewidth}
\hspace*{1.4cm}\mbox{\epsfysize=\hsize\epsffile{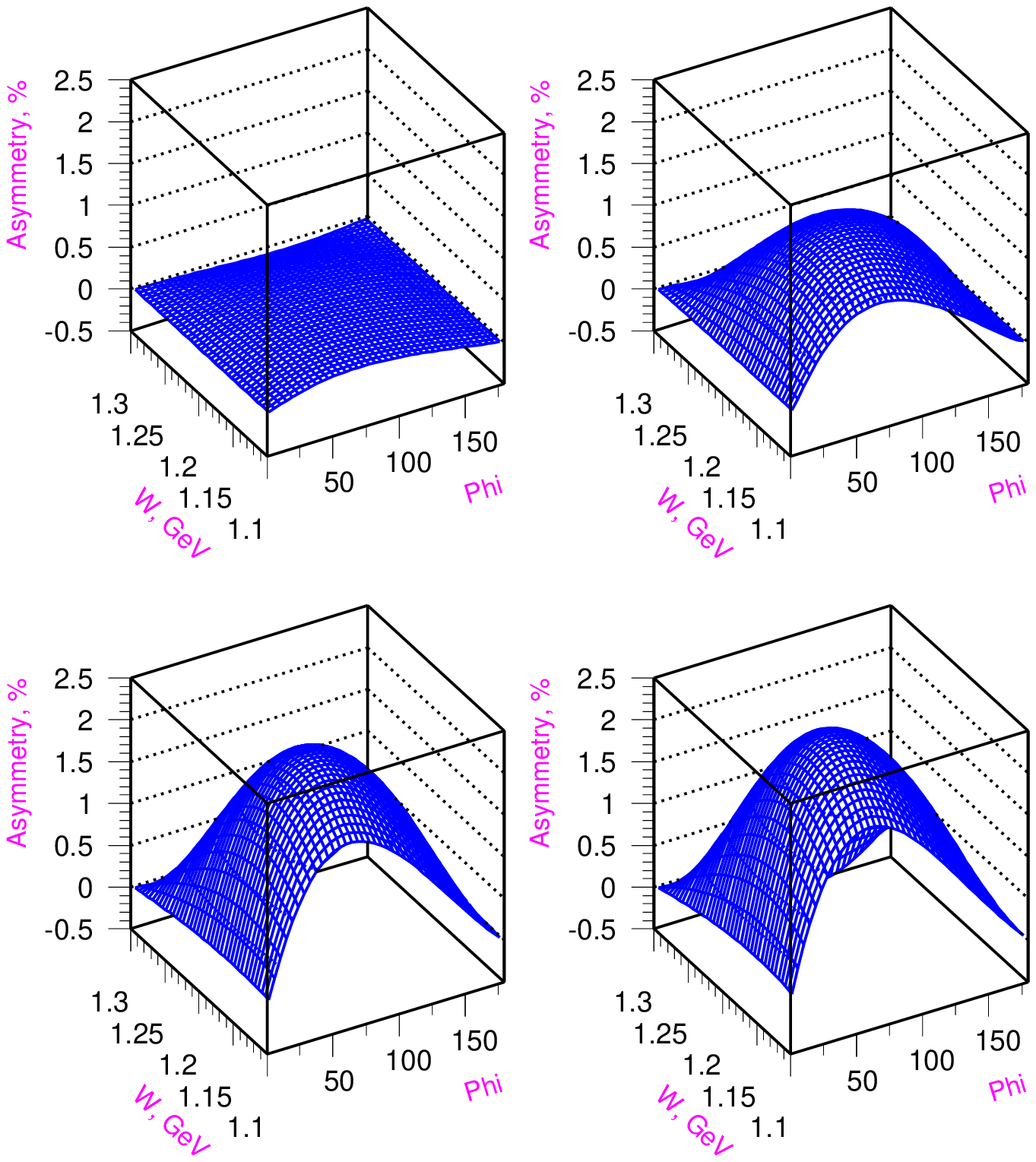}}
 \hfil
   \end{minipage}
\caption{Single spin asymmetry $A_{\rm RC}(\phi,W)$, which 
includes radiative effects evaluated for the values
of the  polar angle $\theta_{\rm cm}$~= 2.5$^\circ$ (upper left),
 13.8$^\circ$ (upper right), 30$^\circ$ (lower left) and
37.5$^\circ$ (lower right).
\label{theta2d}}
\end{figure}
\end{center}

Two dimensional plots for  $A_{\rm {RC}}(\phi, W)$ evaluated by 
employing this approximation are displayed in Fig.~\ref{theta2d}.
\newpage

\begin{figure}[tbh]
\hspace*{2.5cm}\mbox{\epsfysize=0.6\hsize\epsffile{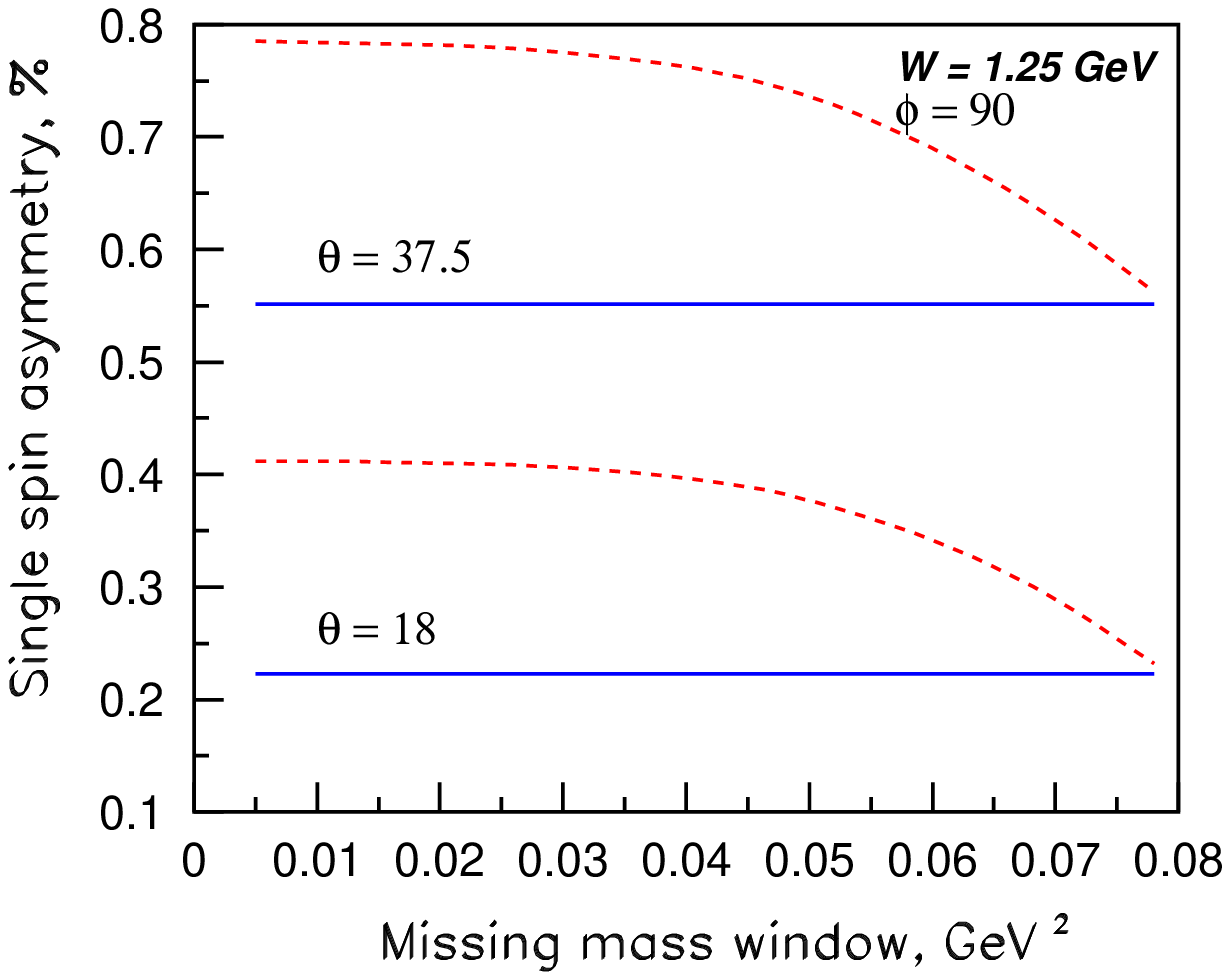}}
\caption{Single spin asymmetry as a function of the $\Delta S$
window evaluated at two data points corresponding to the 
angles $\theta_{\rm cm} =$ 18$^{\circ}$ and 37.5$^{\circ}$. Full and
dashed lines show $A$ and $A_{\rm RC}$, which are 
the Born term and  smeared by radiative effects asymmetries, respectively. 
\label{fig:rc-win-theta}}
\end{figure}

For practical applications the window size $\Delta S$
 was varied in RC calculations between 0.005 and 0.020 GeV$^2$. 
In some plots (e.g. in Fig.~\ref{fig:rc-win-theta}), for illustration
purposes, the window size has been increased up to 0.078 GeV$^2$. 
In the kinematics under study (this time calculations were done at $W= 1.25$ GeV), 
the use of such wide window in the event selection serves an educative 
purpose: it allows to keep virtually all events and obtain the asymmetry, 
which coincides with the Born one.
The considered in Fig.~\ref{fig:rc-win-theta}
 case corresponds to rather large phase space of emission
of a bremsstrahlung photon.
The phase space rapidly contracts when $W$ approaches the lower
boundary of the kinematic region. Therefore, at $W$ = 1.1 GeV
all radiatively distorted events can be found inside rather 
small window of 0.005 GeV$^2$.
 On the contrary, as $W$ approaches 1.3 GeV,
application of the event selection procedure within windows
from 0.005 to 0.024 GeV$^2$, which is a typical size in
the data analysis,  results in a selection of certain
parts of the event samples, which have the following properties
that can be traced in Fig.~\ref{fig:rc-theta-wind}:\\

\begin{tabular}{llll}
(1) ~ ~ & Large $\theta_{\rm cm}$ ~ ~ $\to$& Large RC ~ ~ $\to$ & 
~ ~ No dependence on window size\\
(2) ~ ~ & Small $\theta_{\rm cm}$ ~ ~ $\to$& Small RC ~ ~ $\to$ & 
~ ~ Certain dependence on window size\\
\end{tabular}

Accordingly, when $(\Delta W)^2$ approaches its maximum value,
the corrections vanish, which is seen from converging of curves
in Fig.~\ref{fig:rc-theta-wind} to a single point 
$\delta_{A}$ = 0, and also from convergence of measured and
Born asymmetries displayed in Fig.~\ref{fig:rc-win-theta}.

\section{Summary}
In the kinematics of the considered experiment, the radiative
effects that manifest themselves in the observed single spin 
asymmetry are largest at $W$ = 1.3 GeV and at the $\theta_{\rm cm}$
= 37.5$^{\circ}$. But even there they do not change
the Born asymmetry by more than 1.0 -- 1.5~\% (in absolute value).
A monotonous rise of $A(\theta_{\rm cm})$ and $A_{\rm RC}(\theta_{\rm cm})$ 
in the angular interval (0, 37.5$^{\circ}$) allows to employ simple
analytical approximation of the asymmetries and radiative corrections
by evaluating them in the maximum 
($\theta_{\rm cm}^{\rm max}$ = 37.5$^{\circ}$)
and scaling down to any value of $\theta_{\rm cm}$ inside the 
considered interval. A number of analytical functions, which
approximate numerical results from {\small EXCLURAD}, offer
flexible tools for fast applications of radiative corrections
as a function of $W$, $\theta_{\rm cm}$ and $\phi$.

The dependence of radiative corrections on the  $\Delta S$ window,
if it is chosen as  0.005 $ <\Delta S < $ 0.024 GeV$^2$,
can be neglected in the whole range of $W$ but from the very
different grounds: (1) at  $W$ = 1.1 GeV any  $\Delta S$ window completely 
covers the phase space of radiative events, (2) at $W$ = 1.3 GeV
any $\Delta S$ window is small compared to the phase space of radiative events,
but falls onto the plateau of the dependence of the 
radiative corrections on the  $\Delta S$ window size.
\newpage 

%
%
%
\vspace{-1.cm}
\section{Acknowledgments}
I have benefited from fruitful discussions with H. Fonvieille and 
A.~Afanasev. I am grateful to University of Blaise Pascal
Clermont-Ferrand II for kind hospitality during my work on
this project.

\section{Appendix}
This section explains how the Fortran functions written 
for the approximation of the single spin asymmetries and radiative
corrections can be used in practical work. 

All asymmetries approximated by the functions correspond to
the {\small MAID-2000} parameters used as the input for {\small EXCLURAD}.

The arguments of the functions are in units of ``GeV'' (for the total
c.m. energy $W$) and ``degree'' (for angles $\theta_{\rm cm}$ and
$\phi$).

\begin{enumerate}
\item { {\small FBORNA(PHI,W)}\\
 Returns asymmetry  $A(\phi,W)$ (\%) in Born approximation
at fixed $\theta_{\rm cm}$ = 37.5$^{\circ}$.\\
It has been used in producing left-panel plot of Fig.~\ref{born2d}.}

\item {{\small FRC020(PHI,W)}\\
  Returns asymmetry $A_{\rm RC}(\phi,W)$ (\%), which includes radiative 
effects, at fixed $\theta_{\rm cm}$ = 37.5$^{\circ}$ and corresponds
to the event selection in the missing mass window of 0.02 GeV$^2$.\\
It has been used in producing right-panel plot of Fig.~\ref{born2d} as
well as lower-right plot in Fig.~\ref{theta2d}. }

\item {{\small VRADCOR(PHI,W)}\\
  Returns correction obtained by subtracting  asymmetry 
$A_{\rm RC}(\phi,W)$ from Born asymmetry $A(\phi,W)$.
Uses functions {\small FBORNA(PHI,W)} and {\small FRC020(PHI,W)}.\\
It has been used in producing the plot in Fig.~\ref{diff}.}

\item {{\small FASYMT(PHI,TH)}\\
Returns  $A_{\rm RC}(\phi,\theta)$ (\%) at fixed $W$ = 1.21 GeV
and corresponds to the event selection in the missing 
mass window of 0.02 GeV$^2$.\\
It has been used in producing the plot in Fig.~3.}

\item {{\small FNORM(PHI,TH)}\\
 Returns the ratio of $A_{\rm RC}(\phi,\theta_{\rm cm})$ and
 $A_{\rm RC}(\phi,\theta_{\rm cm}^{\rm max})$ 
for any given $\theta_{\rm cm}$} in the $W$ region around the
mean value of the considered experiment -- $W \approx$ 1.2 GeV.
It has the same structure and uses the same set of fit parameters 
as the function {\small FASYMT(PHI,TH)}.\\
It has been used in producing the plots in Fig.~\ref{theta2d}.
\end{enumerate}


\end{document}